\documentclass[%
 reprint,
 nofootinbib,
 amsmath,
 aps,
pra,
]{revtex4-1}

\usepackage{graphicx}%
\usepackage{dcolumn}%
\usepackage{bm}%
\usepackage[utf8]{inputenc} %
\usepackage[T1]{fontenc}    %
\usepackage{url}            %
\usepackage{booktabs}       %
\usepackage{amsfonts}       %
\usepackage{microtype}      %
\usepackage{amssymb,amsmath,amsthm,graphicx}
\usepackage{xcolor}
\usepackage{comment}
\usepackage{multirow}
\usepackage[caption=false]{subfig}
\usepackage{tabularx}
\usepackage{algpseudocode}
\usepackage{booktabs}
\usepackage{enumitem} 
\usepackage{lipsum}
\usepackage{physics}
\usepackage[pagewise]{lineno}
\usepackage{hyperref}       %

\algrenewcommand\algorithmicrequire{\textbf{Input:}}
\algrenewcommand\algorithmicensure{\textbf{Output:}}

\begin{document}

\title{
Magic state cultivation with lattice-surgery escape on superconducting hardware
}

\author{Haoran Liao}
\email{haoran.liao@q-ctrl.com}
\author{Michael Schutzman}
\author{William Munizzi}
\author{Pranav Mundada}
\author{Gavin Hartnett}
\author{Yuval Baum}
\affiliation{Q-CTRL, Los Angeles, CA USA and Sydney, NSW Australia}

\begin{abstract}
Magic state cultivation (MSC) is a low-overhead protocol for fault-tolerantly preparing logical magic states. 
In this work, we implement two MSC variants to prepare a magic state in a distance-5 code on a publicly available IBM Nighthawk processor. The first is a lattice-surgery escape, recently proposed by Hirano~\textit{et al.} and experimentally studied here for the first time. The second is a grafting escape, previously demonstrated by Rosenfeld~\textit{et al.}, and extended here by incorporating a switch into a matchable code.
Both protocols begin with fault-tolerant cultivation of a $T$ state on a color code, followed by an escape into a larger code.
We decode the results using minimum-weight perfect matching or Tesseract on circuit-level error models calibrated to the hardware noise. Compared with the previous experiment, our cultivated $T$ state achieves a logical error probability of $2\times10^{-5}$ at the $T$-state injection angle, nearly $10$x lower with about $3$x the retainment. Moreover, our execution of the grafting escape has a $2$x lower logical error rate per stabilizer round at a higher retainment. 
When comparing the two protocols under identical execution conditions and complementary gap postselection based on decoder confidence, lattice-surgery escape into the surface code preserves a $T$-state fidelity above $0.99$ ten rounds after escape at about $5\%$ retainment. By contrast, matchable-switched grafting drops below $0.95$ fidelity at comparable retainment. 
Furthermore, the lattice-surgery escape reaches a logical error rate per stabilizer round of about $0.14\%$ with around $11\%$ average shot retainment. 
These results establish lattice-surgery escape as a practical MSC protocol that benefits from a direct transition into a surface code, and demonstrate the lowest MSC error and the highest surface-code magic-state fidelity reported to date, advancing the experimental realization of fault-tolerant non-Clifford gadgets.
\end{abstract}

\maketitle

\begin{figure*}[t]
  \centering
  \includegraphics[width=0.9\textwidth]{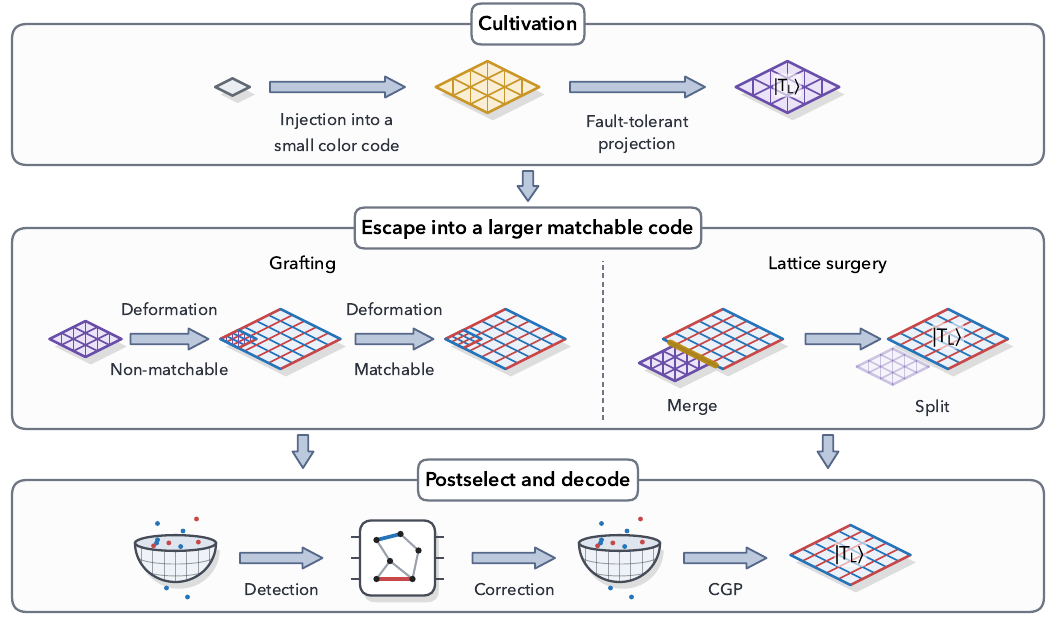}
  \caption{Overview of our experiment workflow. (Top) A physical state is injected into the $[[7,1,3]]$ Steane color code and cultivated into the logical state $\vert{}T_L\rangle$ by fault-tolerant projection. (Middle) The state then escapes into a distance-5 code either by grafting~\cite{Gidney2024Cultivation} (left) with two code deformations through an intermediate non-matchable code to a final matchable code, or by lattice surgery~\cite{Hirano2025MSCLS} (right) with merge and split operations beside an adjacent surface code. (Bottom) Recorded shots are postselected on the error-detecting stabilizer checks and flags, decoded on error-correcting stabilizer checks, and filtered via complementary-gap postselection (CGP).}
  \label{fig:overview}
\end{figure*}

\section{Introduction}
\label{sec:introduction}
Universal fault-tolerant quantum computation (FTQC) requires non-Clifford gates alongside Clifford operations; however, no single quantum error-correcting code can implement a universal gate set transversally~\cite{EastinKnill2009}. In leading architectures such as the surface code~\cite{Kitaev2003,Fowler2012Surface}, Clifford gates are realized fault-tolerantly by transversal operations, code deformation~\cite{Geher2024Hadamard}, and lattice surgery (LS)~\cite{Horsman2012,deBeaudrap2020ZX}, whereas non-Clifford gates are supplied by magic states consumed through fault-tolerant gate teleportation~\cite{GottesmanChuang1999,BravyiKitaev2005}. Among magic states, the $T$ state, which teleports the gate $T=\mathrm{R}_Z(\pi/4)$ from the third level of the Clifford hierarchy~\cite{GottesmanChuang1999}, is the standard target, because Clifford$+T$ is universal~\cite{RossSelinger2016}, and other magic states, such as the $\ket{\mathrm{CC}Z}$ state for the Toffoli gate, can themselves be distilled from $T$ states~\cite{Eastin2013Toffoli}. $T$-state factories account for a substantial fraction of the spacetime resources of many fault-tolerant algorithms~\cite{OGormanCampbell2017}, including those for large-integer factorization~\cite{Gidney2025RSA} and quantum chemistry~\cite{Babbush2018ElectronicSpectra}. Efficient preparation of high-fidelity magic states is therefore essential to making fault-tolerant quantum computation practical.

Two leading fault-tolerant approaches for realizing $T$ gates are magic-state distillation and magic state cultivation (MSC). 
Standard distillation concatenates an outer error-detecting code admitting a transversal $T$ gate with an inner error-correcting code~\cite{BravyiKitaev2005,BravyiHaah2012,Litinski2019Magic}. Transversal $T$ on the outer-code logical $\ket{+}$ yields a noisy outer-code $\ket{T}$, which is postselected on a trivial outer-code syndrome and unencoded through fault-tolerant inner-code gates into a purified inner-code $\ket{T}$.
By contrast, magic state cultivation (MSC) injects a noisy state into a small code and refines it via postselected projections before expanding or deforming it into a higher-distance, matchable code. Operating directly on physical qubits rather than through layers of logical gates substantially reduces spacetime overhead~\cite{Gidney2024Cultivation,Hirano2025MSCLS,Claes2025SurfaceCultivation}.

In the original color-code MSC protocol by Gidney~\textit{et al.}~\cite{Gidney2024Cultivation}, escape into a larger code contributes substantially to logical error and spacetime overhead. Their escape relies on grafting, a code deformation that expands the cultivated color code into an initially non-matchable hybrid color--surface code. Subsequent stabilizer modifications render it matchable, but leave high-weight stabilizers in place.
Hirano~\textit{et al.} proposed MSC with lattice surgery (MSC-LS)~\cite{Hirano2025MSCLS,Chen2026SqrtT}, which teleports the cultivated state directly into a rotated surface code~\cite{Bombin2007Optimal} through a logical $Z_{\text{L}}Z_{\text{L}}$ measurement. MSC-LS provides two key advantages. First, it reduces spacetime overhead at comparable logical error by targeting a more qubit-efficient code with lower-weight stabilizers while yielding higher end-to-end retainment. For instance, in the simulations of Ref.~\cite{Hirano2025MSCLS}, MSC-LS ends in a distance-11 rotated surface code instead of a distance-15 matchable grafted code, roughly halving both the final qubit count and the spacetime volume. Second, using lattice surgery and ending in a surface code simplifies integration into surface-code architectures, avoiding the need for specialized syndrome-extraction circuits and decoders, or downstream code conversion. Other MSC proposals cultivate directly in surface codes using nonlocal gates~\cite{Vaknin2026HighRateMSC,Sahay2025FoldTransversal} and are more suitable for reconfigurable systems.

Logical magic states have been prepared experimentally on trapped-ion~\cite{Daguerre2025}, neutral-atom~\cite{SalesRodriguez2025}, and superconducting~\cite{Gupta2024} processors. To date, however, the only experimental demonstration of MSC was performed by Rosenfeld~\textit{et al.} on Google's Willow processor, where a $T$ state was cultivated in a distance-3 color code and grafted into a distance-5 non-matchable code---leaving the switch to a matchable code unimplemented~\cite{Rosenfeld2025MSC}. 
Meanwhile, the lattice-surgery escape has not been demonstrated experimentally, and no study has yet compared these escape protocols under identical execution (hardware and compilation) conditions.

Here, we present the first experimental realization of MSC-LS and demonstrate that it provides a favorable alternative to grafting for cultivating $T$ states on superconducting hardware. Using a publicly available IBM Nighthawk processor, we teleport a cultivated $T$ state from the Steane code into a distance-5 rotated surface code. 
We optimize the protocol through a streamlined arbitrary-state injection circuit and a minimal Steane-check schedule during the LS merge and split.
On the same device, we execute the grafting-escape protocol of Rosenfeld~\textit{et al.} and 
extend it with a custom circuit enabling a switch into a matchable distance-5 code.
Experimental outcomes are decoded using tailored strategies---minimum-weight perfect matching~\cite{HiggottGidney2025SparseBlossom} (hereafter matching) for our LS circuit, and a hybrid Tesseract~\cite{Beni2025Tesseract,Rosenfeld2025MSC,Sivak2026RLControl} and matching decoder for the matchable-switched grafting---alongside complementary-gap postselection (CGP)~\cite{Gidney2024Cultivation,Bombin2024Postselection,Smith2024Mitigating} to filter shots with low decoder confidence (Secs.~\ref{si:decoding} and~\ref{si:cgp}). Figure~\ref{fig:overview} outlines these stages.

Compared with the results from Rosenfeld~\textit{et al.}~\cite{Rosenfeld2025MSC}, our cultivated $T$ state in the Steane code achieves a nearly order-of-magnitude lower logical error probability at $2 \times 10^{-5}$ (fidelity of at least $0.99998$), with roughly three times the shot retainment. Our execution of the distance-5 non-matchable grafting escape similarly yields higher retainment and fidelity, alongside a reduced logical error rate per stabilizer round. In a direct head-to-head comparison within our executions, the LS escape exceeds a fidelity of $0.99$ under CGP at up to $16\%$ retainment immediately following escape, and sustains this fidelity threshold ten stabilizer rounds after escape at about $5\%$ retainment. In contrast, the matchable-switched grafting exceeds $0.99$ fidelity only up to $11\%$ retainment immediately post-escape, and falls below $0.95$ ten rounds later at comparable retainment. Furthermore, matching decodes MSC-LS over three orders of magnitude faster than the hybrid decoder decodes the matchable-switched grafting. Collectively, these findings establish MSC-LS as a more direct and practical pathway for building $T$-state factories on surface-code fault-tolerant architectures.

\section{Magic state cultivation}
\label{sec:msc}

MSC aims to prepare a magic state efficiently and fault-tolerantly, with a logical error probability far below the physical error rate, in a code that protects it until a gate consumes it~\cite{Gidney2024Cultivation}. Consuming the state, as in a gate teleportation, applies a correction that depends on the decoded errors, so the code must be decoded in real time, keeping pace with its stabilizer rounds. This strongly favors a matchable code, in which every single error flips at most two detectors---decoding then reduces to a matching problem on a graph, which minimum-weight perfect matching solves fast enough~\cite{HiggottGidney2025SparseBlossom}, whereas non-matchable codes need more general and much slower decoders, such as Tesseract~\cite{Beni2025Tesseract}. A magic state, however, is cultivated by measuring a logical operator whose $+1$ eigenstate is the target state and postselecting that outcome. This is most economical in a small code where the operator is transversal, allowing the projection to be implemented efficiently and fault-tolerantly. MSC therefore first cultivates the state in such a code and then lets it escape into a larger, matchable code.

Following Ref.~\cite{Rosenfeld2025MSC}, we target $\ket{T}=T_y\ket{0}$ with $T_y=e^{-i\pi Y/8}$, the $+1$ eigenstate of the Hadamard $H$, which is Clifford-equivalent to the standard $T\ket{+}$; its cultivation thus needs a transversal logical Hadamard operator $H_\text{L}$, as in the $[[7,1,3]]$ Steane color code~\cite{Steane1996,Steane1996Multiple,Perlin2026FTQC}. MSC-LS and grafting, the two escape protocols we compare, both inject a noisy $T$ state into the Steane code and cultivate it by projecting onto the $+1$ eigenstate of $H_\text{L}$ with a circuit that also detects its own faults; rejecting the $-1$ outcome removes logical errors from injection, and rejecting any detected fault prevents the projection itself from introducing one, so the accepted state is prepared fault-tolerantly. As such, cultivation and escape rely mainly on error detection~\cite{Liao2025LongRange}; after escape, the larger code should be error corrected. Since cultivation has fault distance\footnote{The fault distance of a circuit is the minimum number of faults that cause a logical error without flipping any detector~\cite{Hirano2025MSCLS,Chan2026Degraded}.} three, the postselected state's error is $\mathcal{O}(p^3)$ in the physical error rate $p$. Decoding a distance-$d$ code, however, leaves an error per round of $\mathcal{O}(p^{\lceil d/2\rceil})$. Preserving cultivation's cubic error suppression therefore requires an escape code with $d\geq5$. MSC-LS escapes into a distance-5 rotated surface code, whereas grafting escapes into a distance-5 code with up to weight-six stabilizers that becomes matchable only after code switching. 

\begin{figure}[t]
  \centering
  \includegraphics[width=\columnwidth]{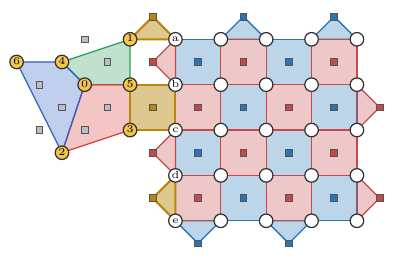}
  \caption{
  Device qubit layout for MSC-LS. The $[[7,1,3]]$ Steane code patch (data qubits 0--6, yellow circles; measure qubits, grey squares) is positioned adjacent to a $[[25,1,5]]$ rotated surface code patch (data qubits, white circles; $X$ and $Z$ stabilizers with their measure qubits, red and blue). The three shaded regions of the Steane code denote its bulk plaquettes, and the gold squares indicate the measure qubits of the three merge checks.}
  \label{fig:ls-layout}
\end{figure}

\begin{figure}[t]
  \centering
  \includegraphics[width=\columnwidth]{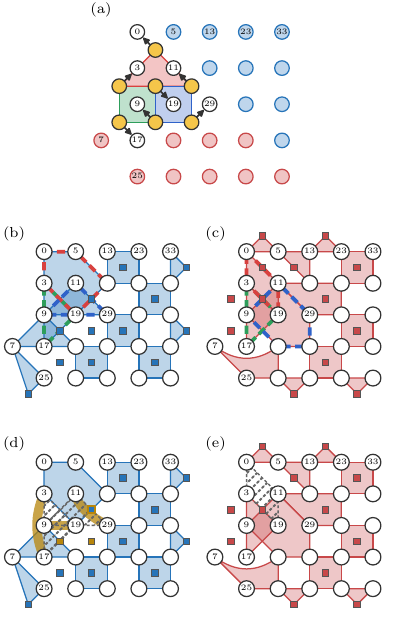}
  \caption{Device qubit layouts of grafting~\cite{Rosenfeld2025MSC}, in (a)--(c), and of the matchable switch, in (d) and (e). (a)~The $[[7,1,3]]$ Steane code, with its plaquettes in green, red, and blue, and where its data move in the extension cycle; the other data qubits of the grafted code are prepared in $\ket{0}$ (blue) or $\ket{+}$ (red). (b), (c)~$Z$ and $X$ stabilizers of the $[[26,1,5]]$ grafted code with their measure qubits (squares); dashed outlines mark the relocated Steane plaquettes, and data qubits named in the text carry their indices. (d), (e)~$Z$ and $X$ stabilizers of the matchable code after the switch; hatched plaquette stabilizers are dropped, two-body $Z$ stabilizers (gold) are added, and the gold-outlined measure qubit reads $Z_{11,29}$.}
  \label{fig:grafting-layouts}
\end{figure}

\subsection{Lattice-surgery escape}
\label{sec:msc-ls}

Our LS implementation follows the protocol of Ref.~\cite{Hirano2025MSCLS} on the 14-site Steane patch of Fig.~\ref{fig:ls-layout}, with a more efficient state-injection circuit (see Sec.~\ref{si:injection} for details). Injection encodes an arbitrary logical state into the Steane code and is followed by one postselected round of Steane checks. These checks, like those during the merge below, use superdense extraction~\cite{GidneyJones2023ColorCode}---two Bell-paired ancillas on each plaquette read out its $X$ and $Z$ stabilizers together (Sec.~\ref{si:superdense}). Cultivation non-destructively measures the logical Hadamard $H_\text{L}=H^{\otimes7}$ once, with a double check~\cite{Rosenfeld2025MSC} and flags (see Sec.~\ref{si:cultivation} for details). Only shots whose double check agrees at $+1$ and whose flags are all trivial are retained.

Lattice surgery then teleports the cultivated state from the Steane code (the source) into a distance-5 rotated surface code (the target) prepared in $\ket{+_{\text{L}}}$ next to the Steane patch: it measures $Z_{\text{L}}Z_{\text{L}}$ of the two logical qubits and then $X_{\text{L}}$ of the source, and the two outcomes determine conditional $X_{\text{L}}$ and $Z_{\text{L}}$ corrections on the target, which we track in the Pauli frame. The $X_{\text{L}}$ measurement is a destructive $X$ readout of the Steane data, which also reconstructs the Steane $X$ stabilizers and so detects its own faults in a single shot. The $Z_{\text{L}}Z_{\text{L}}$ measurement is the product of three merging stabilizers [gold in Fig.~\ref{fig:ls-layout}], $Z_{1a}$, $Z_{35bc}$, and $Z_{de}$, where $P_f=\prod_{j\in f}P_j$ for a Pauli $P$ on the qubits in $f$, which join Steane data 1, 3, and 5 to the surface code's boundary data $a,\dots,e$. This measurement has no such redundancy---one measurement error would flip its outcome undetected, so it is repeated $\mathcal{O}(d)$ times for fault distance $d$, in three rounds here. Both codes must stay stabilized during these merge rounds by measuring stabilizers that commute with the $Z_{\text{L}}Z_{\text{L}}$, as detailed in Ref.~\cite{Hirano2025MSCLS} and Sec.~\ref{si:escape}. Unlike Ref.~\cite{Hirano2025MSCLS}, we perform only a minimal subset of the Steane $Z$ checks during the merge rounds (see Sec.~\ref{si:escape} for details); together with the destructive $X$ readout of the Steane data, this subset still preserves fault distance three. We establish this and the other fault distances below by a complete search in fault space~\cite{Vaknin2026HighRateMSC}, enumerating every single fault and pair of faults. For a non-Clifford circuit, we run this search on a Clifford proxy with the $S$ state in place of the $T$ state. We neglect the resulting $T$--$S$ discrepancy, which may overestimate the fault distance of the non-Clifford circuit and hence its error suppression at low physical error rates~\cite{Chan2026Degraded,WanZapirain2026Exact}. After lattice surgery, the surface code resumes full stabilizer rounds. Steane checks and their flags, and changes of the merging stabilizers between rounds, are postselected; the surface-code checks are decoded. Our demonstration stops at this distance-5 surface code, without the subsequent expansion to a larger distance in the full MSC-LS protocol~\cite{Hirano2025MSCLS}.

\subsection{Grafting escape and matchable switch}
\label{sec:grafting}

We adopt the grafting circuit of Ref.~\cite{Rosenfeld2025MSC} with one addition: a stabilizer round after injection, without which the circuit has fault distance only two. Its pre-escape circuit, injection and cultivation in the Steane code, is similar to that of our LS circuit. Grafting then expands the Steane code in one extension cycle---the Steane data move outward on the square-lattice grid to enable better layout arrangement in the larger code [Fig.~\ref{fig:grafting-layouts}(a)], fresh data qubits are prepared around them, and the cycle measures every stabilizer of a $[[26,1,5]]$ grafted code [Fig.~\ref{fig:grafting-layouts}(b,c)]. Each fresh data qubit is prepared in $\ket{0}$ or $\ket{+}$. Each grafted-code logical operator is a Steane logical operator multiplied by $X$ or $Z$ stabilizers of the fresh qubits; $X_{\text{L}}$ also includes a Steane stabilizer whose value is known from its last readout. Explicitly, in the qubit indices of Fig.~\ref{fig:grafting-layouts}(b,c), where the relocated Steane code has the logical representatives $X_{\text{L}}=X_{0,9,19}$ and $Z_{\text{L}}=Z_{0,3,17}$, the grafted logical operators can be represented as
\begin{align}
  X_{\text{L}}' &= X_{\text{L}}\,X_{3,9,17,19}\,X_{7,25} = X_{0,3,7,17,25},\nonumber\\
  Z_{\text{L}}' &= Z_{\text{L}}\,Z_{5,13,23,33} = Z_{0,3,5,13,17,23,33},
  \label{eq:grafted-logicals}
\end{align}
where $X_{3,9,17,19}$ is an $X$ stabilizer of the original Steane code, qubits 7 and 25 are prepared in $\ket{+}$, and 5, 13, 23, 33 in $\ket{0}$. These products commute with every measured stabilizer of the grafted code, so the code deformation preserves the logical state; their known signs enter the observable parity. The three Steane plaquettes relocate to the corner, where two of them grow to weight six. A data qubit there lies in three stabilizers of one type, so the code is not matchable.

To switch into a matchable code, we adapt the stabilizer configuration from Ref.~\cite{Gidney2024Cultivation}. The $Z$ stabilizers of the green and blue plaquettes, $Z_{3,9,17,19}$ and $Z_{9,11,19,29}$, and the $X$ stabilizer of the red one, $X_{0,3,11,19}$, are dropped, and the two-body stabilizers $Z_{3,17}$, $Z_{9,19}$, and $Z_{11,29}$ are added, whose products recover the dropped $Z$ stabilizers [Fig.~\ref{fig:grafting-layouts}(d,e)]. The added stabilizers commute with $X_{\text{L}}'$ and $Z_{\text{L}}'$ of Eq.~\eqref{eq:grafted-logicals} and with all remaining stabilizers, so measuring them preserves the logical state. Every data qubit then lies in at most two stabilizers of each type, so the code is matchable, though it retains the high-weight corner stabilizers. We adopt this deformation because it maintains the output's protection at code distance five with limited changes to the non-matchable grafted-code stabilizer measurement circuit.

The matchable switch that ablates the color-code region must not come too early, because the non-matchable grafted code must first be stabilized. Heuristically, a stabilizer about to be dropped must first be measured in $d-1$ cycles to reach fault distance $d$, because a data error can evade each cycle's detector only with one more measurement error. Exhaustive single- and pair-fault search confirms that two non-matchable grafted cycles (totaling three cycles including the extension), which we use, preserve fault distance three, the limit set by pre-escape cultivation, for both logical readouts. Since MSC aims to end in a matchable code, we count the extension cycle, these two non-matchable grafted cycles, and the matchable switch cycle together as the escape of this protocol, which we call the matchable-switched grafting.

\section{Results}
\label{sec:results}

We are concerned with whether MSC-LS is a better way than grafting to cultivate $T$ states for a $T$-state factory on superconducting hardware. Such a factory must prepare a high-fidelity state, move it into a larger code efficiently, and preserve it there, with fast enough decoding, until it is consumed. We follow the state through these stages in turn in experiment. We first characterize the state cultivated in the Steane code, from which every escape starts (Sec.~\ref{sec:results-kt}). We then execute side by side on the same device the two escapes that end in a matchable code, MSC-LS and the matchable-switched grafting, and, as a reference, grafting without the switch, so that their differences reflect the advantages and disadvantages of ending in a matchable rather than a non-matchable code. We compare how much fidelity and how many shots each escape costs (Sec.~\ref{sec:results-immediate}), how well each final code preserves the state (Sec.~\ref{sec:results-storage}), and how fast each whole protocol is decoded in postprocessing (Sec.~\ref{si:decoding}).

\subsection{Cultivated \texorpdfstring{$T$}{T} state}
\label{sec:results-kt}

\begin{figure}[t]
  \centering
  \includegraphics[width=\columnwidth]{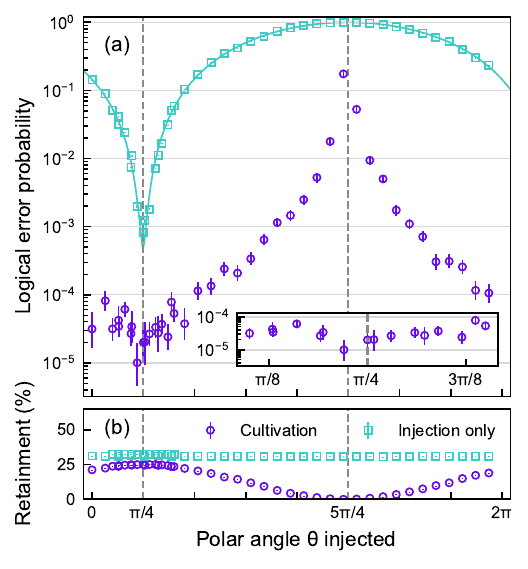}
  \caption{$T$-state quality metrics based on kickback tomography. (a)~Logical error probability and (b)~shot retainment as a function of the injected polar angle $\theta$, comparing cultivation (circles) to state injection without cultivation (squares). Error bars denote 68\% confidence intervals, and the solid curve is a fit to the injection-only logical error probability. The inset provides an expanded view of the region $0.1\pi\leq\theta\leq0.4\pi$. Dashed vertical lines indicate injection angles for the $T$ state and its orthogonal state.}
  \label{fig:kt}
\end{figure}

We first characterize the cultivated state in the pre-escape stage on \texttt{ibm\_phoenix} with kickback tomography (KT), whose construction is detailed in Ref.~\cite{Rosenfeld2025MSC}. After injection, one cultivation round, named the preparation round, cultivates the state, and two further cultivation rounds measure it along the $T$ axis, optionally interleaved with Steane checks. A shot is retained when every detector is quiet and the repeated projections agree. The KT logical error is the fraction of retained shots that do not measure the $T$ state, an upper bound on the infidelity of the cultivated state~\cite{Rosenfeld2025MSC}. We vary the injection polar angle $\theta$ of the input $\ket{\theta}=e^{-i\theta Y/2}\ket{0}$, which sweeps the $XZ$ great circle of the Bloch sphere with $\theta=\pi/4$ giving $\ket{T}$, and repeat each angle with an injection-only reference that omits the preparation round. Each of the 11 angles from $0.1\pi$ to $0.4\pi$ has $1{,}200{,}000$ shots, and each other angle has $600{,}000$ (Fig.~\ref{fig:kt} and Sec.~\ref{si:kt}).

Our execution produces a $T$ state with a fidelity of at least $0.99998$. This corresponds to an almost order-of-magnitude lower logical error probability than in Ref.~\cite{Rosenfeld2025MSC}, at about three times the retainment. At the $T$-state injection angle $\theta=\pi/4$, the KT logical error is $2.0\,[1.2,3.2]\times10^{-5}$ (6 errors in $300{,}462$ retained shots) at a retainment of $25.0(8)\%$, and over the 11 angles from $0.1\pi$ to $0.4\pi$ it is $3.2(3)\times10^{-5}$ (Fig.~\ref{fig:kt}). At $\theta=\pi/4$, cultivation suppresses the injection-only logical error by a factor of 41. Our injection and cultivation circuits differ only slightly and insignificantly from those of Ref.~\cite{Rosenfeld2025MSC} (see Sec.~\ref{si:kt}). This comparison suggests that much of the improvement in this characterization comes from the latest hardware. The KT results support treating pre-escape preparation error as small compared with the observed escape errors, providing a sound starting point for a meaningful demonstration of MSC-LS and comparison of the escape protocols.

\begin{figure*}[t]
  \centering
  \includegraphics{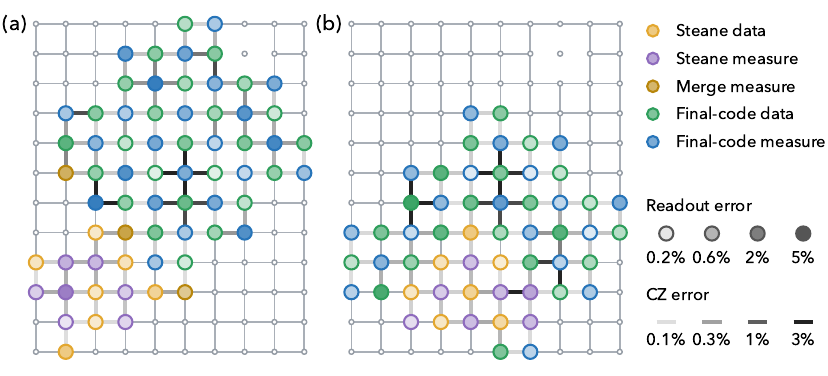}
  \caption{Qubit placement of (a)~MSC-LS and (b)~grafting on \texttt{ibm\_phoenix}, colored by role; the final code is the surface code in (a) and the grafted code in (b), and in (b) qubits used before escape carry their Steane role. Small open circles and thin grey lines are unused qubits and couplers. The matchable-switched grafting uses the same placement as grafting.}
  \label{fig:device}
\end{figure*}

\subsection{Escape}
\label{sec:results-escape}

We compare three escape circuits, our LS circuit, the grafting circuit, and the matchable-switched grafting, each as a memory sweep over $N=2,\dots,10$. Here, $N$ is the number of stabilizer rounds of the final code, which for grafting are the extension cycle and the grafted cycles that follow it, and for the matchable-switched grafting are the matchable cycles after the switch cycle. Each circuit has $500{,}000$ shots per tomography basis and $N$. All escape circuits use the same layout-selection procedure (Sec.~\ref{si:layout}) and transpilation, but since LS uses 66 qubits and grafting, with or without the switch, 55, the procedure places their Steane codes on different qubits (Fig.~\ref{fig:device}). All use dynamical decoupling~\cite{Coote2025GraphDD} and 25 Pauli twirls per circuit, and they were all executed in a single batch, so they share one device state. We decode MSC-LS with matching, grafting with Tesseract, and the matchable-switched grafting with the hybrid decoder, each on a circuit-level error model from the device calibration (Sec.~\ref{si:decoding}).

From the postselected and decoded logical $X$ and $Z$ expectations after $N$ rounds, we estimate the fidelity with $\ket{T}$, $F=[1+(\langle X_{\text{L}}\rangle+\langle Z_{\text{L}}\rangle)/\sqrt{2}]/2$. Retainment is the fraction of all shots that pass postselection and, where applied, the CGP threshold, which discards the shots whose decoder is not confident of its logical prediction (Sec.~\ref{si:cgp}). We compare the escapes in two regimes. Immediately after escape, we compare the fidelity and retainment at $N=3$ against the CGP threshold, which shows how much fidelity and how many shots each escape costs. We take $N=3$ rather than $N=5$ because the escaped state inherits fault distance three from cultivation in the Steane code. In memory after escape, we compare how well each final code preserves the state, through the highest observed fidelity at each $N$ up to $N=10$ and the logical error rate per stabilizer round $p_{\text{L}}$ against the CGP threshold. As in Ref.~\cite{Rosenfeld2025MSC}, we estimate $p_{\text{L}}$ from a fit of the $XZ$-plane Bloch radius $R(N)=(\langle X_{\text{L}}\rangle^2+\langle Z_{\text{L}}\rangle^2)^{1/2}$ over $N=3,\dots,10$, from immediately after escape, to $A\,e^{-2p_{\text{L}}N}$ (Sec.~\ref{si:escape-results}), with $A$ the amplitude. For the matchable-switched grafting, the fit runs over $N=4,\dots,10$, because up to $N=3$ its hybrid decoder decodes the whole history with Tesseract, and only from $N=4$ does it decode the remaining rounds with matching (Sec.~\ref{si:decoding}).

\subsubsection{Immediately after escape}
\label{sec:results-immediate}

MSC-LS and grafting use similar injection and cultivation circuits in the Steane code, so differences after escape mainly reflect the escape methods and final codes. Before escape, LS retains more shots than the matchable-switched grafting, $70.2\%$ against $60.9\%$ (Table~\ref{tab:stages}), largely because the patch layouts place the two Steane codes on qubits with unequal error rates (Sec.~\ref{si:stages}). The escape reverses this order. Grafting escapes in a single extension cycle that postselects six detectors from Steane checks, whereas the MSC-LS escape postselects 20, from the repeated merge checks, the Steane checks during the merge and at destructive readout, and flags. Without CGP, LS therefore retains $30.1\%$ of all shots at $N=3$ against $43.5\%$ for the matchable-switched grafting, and grafting about as many as the latter, since the switch adds no postselection.

\begin{figure*}[t]
  \centering
  \includegraphics[width=\textwidth]{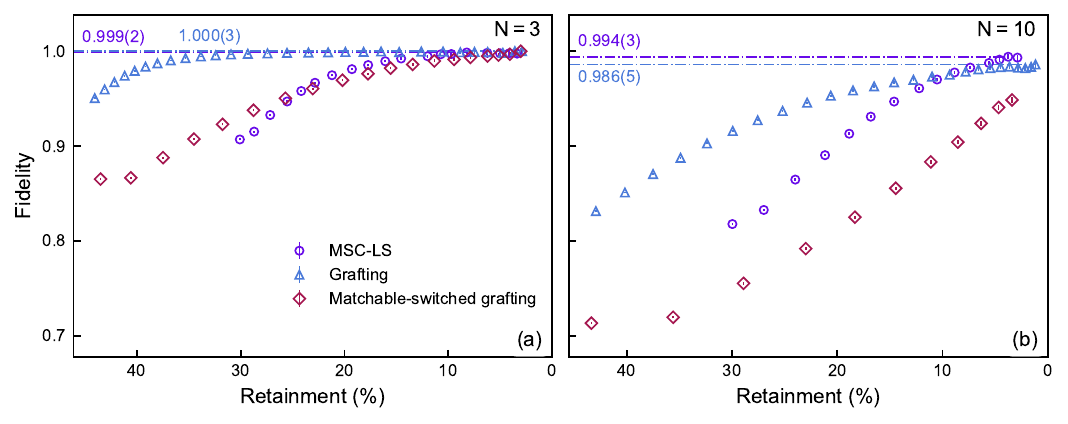}
  \caption{Fidelity with $\ket{T}$ against the retainment at (a)~$N=3$ (left) and (b)~$N=10$ (right), from no CGP (leftmost) through rising CGP thresholds, on a common fidelity scale, for MSC-LS decoded with matching (circles), grafting decoded with Tesseract (triangles), and the matchable-switched grafting decoded with the hybrid decoder (diamonds). Dash-dot lines mark the highest MSC-LS and grafting fidelities, labeled with their values; error bars are one standard error.}
  \label{fig:fidelity}
\end{figure*}

Immediately after escape, grafting has the highest fidelity at a given retainment (Fig.~\ref{fig:fidelity}(a)); without CGP, it is followed by LS and then the matchable-switched grafting. Our grafting also exceeds that on the Willow processor~\cite{Rosenfeld2025MSC} in both fidelity and retainment (Sec.~\ref{si:willow}). With CGP, grafting and LS both exceed a fidelity of $0.99$, grafting at retainments up to $36.7\%$ and LS up to $15.8\%$, and both reach about $0.999$ at the highest thresholds, whereas the matchable-switched grafting exceeds $0.99$ only up to $10.9\%$ (Sec.~\ref{si:cgp}).

These differences follow from the escape circuits. The single extension cycle of grafting is shallower than the three merge rounds and the destructive readout of the LS escape (Table~\ref{tab:stage-depths}), so it suffers less decoherence~\cite{Liao2023DecoheringTN}, and its higher retainment lets it apply a stricter CGP threshold at the same retainment. The matchable-switched grafting shares this escape, but at the same $N$ it has also passed through the two grafted cycles and the switch cycle, whose errors lower its fidelity. Grafting without the switch thus delivers a high-fidelity state at the lowest cost in shots. It is, however, hardly a fair reference for a $T$-state factory, because it ends in a non-matchable code, whose decoding with Tesseract is more than three orders of magnitude slower than matching (Sec.~\ref{si:decoding-cost}) and is expected to be incompatible with the real-time decoding on superconducting hardware~\cite{Acharya2025BelowThreshold} needed to preserve a state until it is consumed. Among the escapes that end in a matchable code, MSC-LS wins immediately after escape, retaining more shots at every fidelity above about $0.95$.

\subsubsection{Memory after escape}
\label{sec:results-storage}

\begin{figure}[t]
  \centering
  \includegraphics[width=\columnwidth]{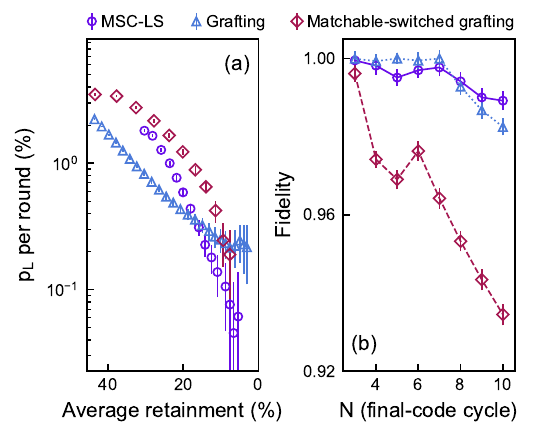}
  \caption{(a)~Logical error rate per stabilizer round $p_{\text{L}}$ against the average retainment over both tomography bases and the fitted $N$, from no CGP (leftmost) through rising CGP thresholds, with 95\% bootstrap intervals, and (b)~highest observed fidelity with $\ket{T}$ over CGP thresholds that retain at least $5\%$ of the shots at each $N$, with one standard error, for MSC-LS decoded with matching (circles), grafting decoded with Tesseract (triangles), and the matchable-switched grafting decoded with the hybrid decoder (diamonds).}
  \label{fig:memory}
\end{figure}

Without CGP, LS has the lowest $p_{\text{L}}$, followed by grafting and then the matchable-switched grafting (Fig.~\ref{fig:memory}(a)). Our grafting $p_{\text{L}}$ is less than half the $5.3\%$ on the Willow processor~\cite{Rosenfeld2025MSC}, decoded and fitted alike, at a higher retainment (Sec.~\ref{si:willow}). With CGP, LS reaches $0.14(5)\%$ at about $11\%$ average retainment, whereas grafting stops falling near $0.22\%$, partly because Tesseract's complementary gap is limited by its already-expensive search budget on shots with many detection events, where matching's remains informative (Sec.~\ref{si:cgp}). The matchable-switched grafting approaches LS only at lower retainment (Sec.~\ref{si:comparisons}). At sufficiently strong postselection, LS thus has the lowest logical error rate per round.

By $N=10$, the highest fidelity favors LS (Fig.~\ref{fig:memory}(b)). The highest observed fidelities of LS and grafting stay close up to $N=7$, grafting slightly ahead, after which grafting falls increasingly below LS, as the lower $p_{\text{L}}$ of LS outweighs the fidelity it loses in its deeper escape. At $N=10$, only LS exceeds a fidelity of $0.99$ at retainments up to $4.6\%$. Among the points plotted in Fig.~\ref{fig:fidelity}(b), LS reaches its highest fidelity of $0.994(3)$ at $3.8\%$ retainment, compared with $0.986(5)$ at $1.2\%$ for grafting. Since LS also has the shallower final-code cycle (Table~\ref{tab:stage-depths}), this gap likely widens at larger $N$. Grafting's retainment advantage at high fidelity immediately after escape thus disappears in memory.

The matchable-switched grafting also ends in a matchable code, but its fidelity stays the lowest at every $N$ and below $0.95$ at $N=10$, because the grafted cycles and the switch before its matchable code have already lowered it; its rise from $N=5$ to $6$ is most likely a statistical fluctuation (Sec.~\ref{si:comparisons}). Those non-matchable grafted cycles also still require Tesseract, so its decoding remains far slower than matching on LS (Sec.~\ref{si:decoding-cost}). Among the protocols, MSC-LS thus combines high fidelity with fast decoding by escaping directly into a matchable code.

\section{Conclusion and outlook}
\label{sec:conclusion}

We report the first experimental demonstration of magic state cultivation with lattice-surgery escape (MSC-LS) on superconducting hardware. We cultivated a Steane-code $T$ state to a kickback-tomography logical error probability of $2\times 10^{-5}$ at $25\%$ retainment before teleporting it into a distance-5 rotated surface code. Under matched conditions, we also implemented and extended the grafting escape of Rosenfeld~\textit{et al.}~\cite{Rosenfeld2025MSC} with a switch to a matchable code. Our MSC-LS state achieves the highest fidelity reported to date for a magic state in a surface code, exceeding $0.99$ at up to $16\%$ retainment immediately after escape and maintaining $0.991(2)$ at $4.6\%$ retainment after ten stabilizer rounds. By comparison, non-fault-tolerant preparations in distance-3 surface codes reached $0.9090(9)$ at $73\%$ retainment on Zuchongzhi~2.1~\cite{Ye2023MagicBeyond}, $0.8806(2)$ at $36\%$ on \texttt{ibm\_fez}~\cite{Kim2026IBMInjection}, and $0.814$ at $2.3\%$ after teleportation using three such code patches on \texttt{ibm\_phoenix}~\cite{Harper2026Entangled}. Grafting into a distance-5 non-matchable code on Willow~\cite{Rosenfeld2025MSC} also yielded lower fidelity (Sec.~\ref{sec:results-immediate}).

Within our experiment, the comparison between MSC-LS and grafting reveals two regimes. Immediately post-escape, grafting without the matchable switch benefits from a single shallow extension cycle that postselects on fewer detectors, yielding higher initial retainment and fidelity. In the memory phase, MSC-LS proves superior: matching with CGP achieves the lowest per-round logical error rate ($\sim 0.14\%$), and MSC-LS is the only protocol to sustain fidelity above $0.99$ after ten stabilizer rounds. In contrast, although the matchable-switched grafting ends in a matchable code, the error accumulated across the preceding grafting and switch cycles depresses its ten-round fidelity below $0.95$ at comparable retainment. Crucially, matching decodes MSC-LS over three orders of magnitude faster than Tesseract-based decoders decode the grafted escapes. Given that real-world $T$-state factories demand both robust memory until consumption and decoding in real time, these findings favor lattice-surgery escape for scalable surface-code architectures.

These results also carry broader implications for fault-tolerant gadgets beyond MSC. A code deformation must be evaluated across its entire trajectory, rather than solely by its target code. Although the matchable-switched grafting ends in a matchable code, the preceding non-matchable cycles degrade logical fidelity and still mandate slow decoding. State transfer between codes therefore benefits from entering a matchable code directly. Furthermore, the choice of decoder affects the effectiveness of CGP. Matching's complementary gap remains informative even for shots with many detection events on contemporary hardware, whereas Tesseract's gap estimation is constrained by its practical search budget (Sec.~\ref{si:cgp}). As a result, the decoding requirements imposed by the full protocol can limit how far postselection can suppress logical errors. Finally, lattice-surgery escape uses the same operations and decoding primitives as the surrounding surface-code architecture (as also noted in Ref.~\cite{Hirano2025MSCLS}), streamlining the overall FTQC stack.

Several avenues remain open for future work. First, our target code terminates at distance 5, whereas the complete MSC-LS protocol expands the surface code to larger distances, such as the distance-11 target analyzed numerically in Ref.~\cite{Hirano2025MSCLS}, whose experimental realization demands a substantially larger device. Second, in the absence of CGP, numerical simulations based on calibrated Pauli-noise models and Clifford-proxy cultivation correctly reproduce the relative performance hierarchy of the two escape methods but notably underestimate their absolute logical error rates per stabilizer round (Sec.~\ref{si:simulation}). This discrepancy motivates further investigation of the Clifford approximation and unmodeled noise, including possible contributions from leakage, crosstalk, and coherent errors. Finally, consuming the cultivated state to teleport a logical $T$ gate into a broader fault-tolerant algorithm necessitates low-latency decoding and real-time feedforward of the required corrections. Integrating these capabilities remains a demanding engineering challenge beyond the offline decoding used here.

\subsection*{Author contributions}
H.L. initiated the project, optimized and extended the protocols, and conducted the simulation. H.L. and M.S. performed the experiments and analysis. All authors contributed to the writing and review of the manuscript.

\subsection*{Data availability}
The dataset supporting the findings of this study, including measurement counts, quantum circuits, and corresponding device calibration data, is available on Zenodo at \url{https://doi.org/10.5281/zenodo.24098457}.

\subsection*{Acknowledgments}
We would like to thank Yutaka Hirano for helpful discussions of the MSC-LS protocol. We are grateful to all other colleagues at Q-CTRL whose technical work has supported the results presented in this paper. We acknowledge the use of IBM Quantum Credits via the IBM Quantum Startups Program for this work. The views expressed are those of the authors and do not reflect the official policy or position of IBM or the IBM Quantum Platform team.

\clearpage
\setcounter{section}{0}
\setcounter{equation}{0}
\setcounter{figure}{0}
\setcounter{table}{0}
\renewcommand{\thesection}{S\arabic{section}}
\renewcommand{\theequation}{S\arabic{equation}}
\renewcommand{\thefigure}{S\arabic{figure}}
\renewcommand{\thetable}{S\Roman{table}}
\renewcommand{\theHsection}{S\arabic{section}}
\renewcommand{\theHequation}{S\arabic{equation}}
\renewcommand{\theHfigure}{S\arabic{figure}}
\renewcommand{\theHtable}{S\arabic{table}}
\begin{center}
  \textbf{Supplementary Information}
\end{center}

\section{Injection}
\label{si:injection}

\begin{figure}[t]
  \centering
  \includegraphics[width=0.4\columnwidth]{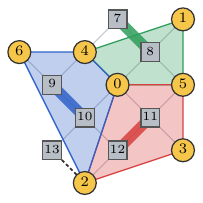}
  \caption{The 14-site Steane patch of Fig.~\ref{fig:ls-layout}, with data qubits 0--6 (yellow circles) and ancillas 7--13 (grey squares), joined by grey lines where they couple. The three colors indicate the three plaquettes, and a band in each plaquette's color joins its Bell pair of ancillas, the $X$ ancilla 7, 11, or 9 and the $Z$ ancilla 8, 12, or 10. The dashed line joins data qubit 2 to its copy on site 13.}
  \label{fig:steane-patch}
\end{figure}

We label the patch sites as in Fig.~\ref{fig:steane-patch}. The Steane code is a self-dual CSS code, so its $X$ and $Z$ stabilizers share the same set of supports, generated under $\oplus$ by $g_u=0246$, $g_v=1256$, and $g_w=0136$, and its logical operators are $X_{\text{L}}=X_\ell$ and $Z_{\text{L}}=Z_\ell$ with $\ell=0123456$. Our unitary encoder, which is not fault-tolerant, takes an arbitrary state $\ket{\psi}$ on data qubit 3 to $\ket{\psi_{\text{L}}}$ on qubits $0,\dots,6$, with all other patch sites initially in $\ket{0}$. It uses 23 nearest-neighbor C$X$ gates in nine layers on 12 of the 14 patch sites, against 25 C$X$ gates in 11 layers in Ref.~\cite{Hirano2025MSCLS}, and returns all non-data qubits disentangled to $\ket{0}$. The non-data qubits could serve as flags, though we chose not to use them since the injection circuit is shallow. We construct it in four steps.

First, the encoding map follows from the stabilizers. Let $S$ be this set of supports, as binary vectors. A computational-basis input $\ket{a}$ is encoded as
\begin{equation}
  \begin{aligned}
    \ket{a}\mapsto\ket{a_{\text{L}}}&=2^{-3/2}\sum_{s\in S}\ket{a\ell\oplus s}\\
    &=2^{-3/2}\sum_{u,v,w\in\{0,1\}}\ket{a\ell\oplus ug_u\oplus vg_v\oplus wg_w}.
  \end{aligned}
  \label{eq:encoding-map}
\end{equation}
Every word in the sum overlaps each stabilizer support evenly, so all $Z$ stabilizers are $+1$, and each $X$ stabilizer adds an element of $S$ to every word and thus permutes the terms. By linearity, the map encodes an arbitrary superposition of $\ket{0}$ and $\ket{1}$. The sum over the bits $u,v,w$ is supplied coherently by data qubits 2, 0, and 5 prepared in $\ket{+}$.

\begin{figure}[t]
  \centering
  \includegraphics[width=\columnwidth]{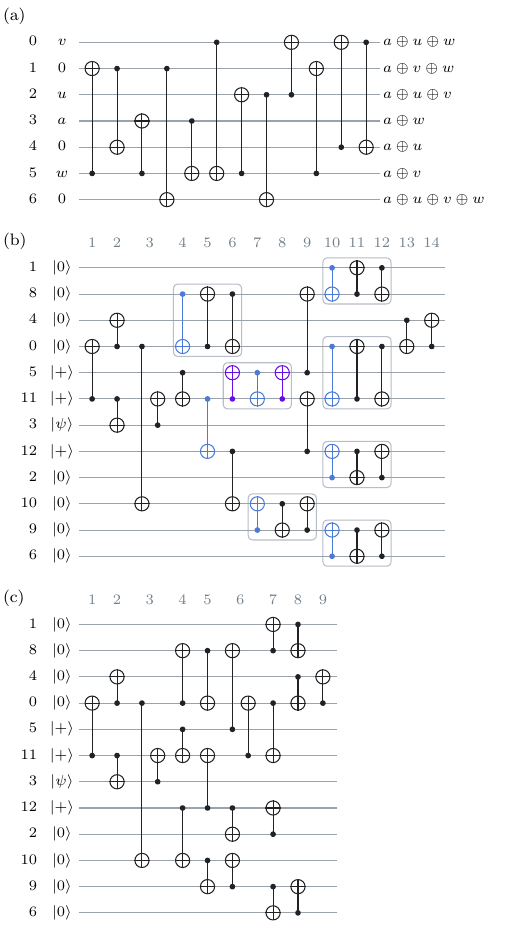}
  \caption{Construction of the encoder. (a)~The XOR network on seven wires, numbered by the data qubit that receives their sum, with the wire bits on the left and the sums on the right. (b)~The network routed on the patch sites of Fig.~\ref{fig:steane-patch}, with their initial states on the left and each SWAP expanded into three C$X$ gates (grey boxes). The blue C$X$ gates are removed, after which the two purple C$X$ gates cancel. (c)~The remaining 23 C$X$ gates in their nine layers.}
  \label{fig:encoder-construction}
\end{figure}

Second, since a C$X$ adds its control bit to its target bit, the encoder is a reversible XOR network that turns the wires $(v,0,u,a,0,w,0)$ into these seven sums [Fig.~\ref{fig:encoder-construction}(a)].

Third, we place the seven wires on the patch and route the network with nearest-neighbor C$X$ gates and seven SWAPs, which bring each sum to its data qubit. Expanding each SWAP into three C$X$ gates gives 33 C$X$ gates [Fig.~\ref{fig:encoder-construction}(b)].

Fourth, we remove gates that act trivially on the known initial states of the sites other than the input qubit, since a C$X$ whose control is in $\ket{0}$ or whose target is in $\ket{+}$ acts as the identity. This reduces six SWAPs that exchange a state with $\ket{0}$ to two-C$X$ moves, removes the middle C$X$ of a seventh SWAP, and deletes one C$X$ of the XOR network, leaving 25 C$X$ gates [blue in Fig.~\ref{fig:encoder-construction}(b)]. Two equal C$X$ gates (purple) then cancel, leaving 23, and scheduling each gate in its earliest layer gives nine layers [Fig.~\ref{fig:encoder-construction}(c)]. 

\section{Superdense stabilizer extraction}
\label{si:superdense}

Our superdense extraction~\cite{GidneyJones2023ColorCode} (Sec.~\ref{sec:msc-ls}) is executed on the patch of Fig.~\ref{fig:steane-patch}~\cite{Hirano2025MSCLS}. It relies on a Bell pair of ancillas being a simultaneous $+1$ eigenstate of the commuting $XX$ and $ZZ$. Coupling the data of a plaquette $f$ to its pair makes the joint state a $+1$ eigenstate of $XX\,X_f$ and $ZZ\,Z_f$. These still commute when $f$ has even weight, which holds here because every plaquette has weight-4, so a Bell measurement of the pair returns $X_f$ and $Z_f$ together. As in superdense coding, one pair carries two bits.

Each plaquette has its own Bell pair of neighboring ancillas (Fig.~\ref{fig:steane-patch}). Data qubit 2 neighbors the ancillas of only one of its two plaquettes, so in each basis in which the other plaquette is measured, it is temporarily copied onto site 13, which stands in for it. For the $Z$ stabilizer, site 13 starts in $\ket{0}$, and a C$X$ from qubit 2 to site 13 makes the two a repetition code with stabilizer $Z_{2,13}$, so multiplying $Z_{0246}$ by it moves its support from qubit 2 to site 13. For the $X$ stabilizer, site 13 starts in $\ket{+}$, and a C$X$ from site 13 to qubit 2 gives the stabilizer $X_{2,13}$ instead. Repeating the C$X$ undoes each copy. When only one basis of a plaquette is measured, the readout of its unused ancilla becomes a flag, and the final readout of site 13 is also a flag. A round of all six stabilizers uses 34 C$X$ gates in 12 layers and seven readouts.

\section{Cultivation}
\label{si:cultivation}

We use one preparation round, postselecting the $+1$ outcome of the $H_\text{L}$ measurement~\cite{Rosenfeld2025MSC, Hirano2025MSCLS}. With transversal $H_{\text{L}}=H^{\otimes7}$, its projector is implemented as
\begin{equation}
  \frac{I+H_{\text{L}}}{2}=T_y^{\dagger\otimes7}\,\frac{I+X^{\otimes7}}{2}\,T_y^{\otimes7},
  \label{eq:hadamard-projector}
\end{equation}
where $T_y=e^{-i\pi Y/8}$ rotates the Hadamard axis onto the $X$ axis. Between transversal $T_y$ and $T_y^\dagger$ layers, the $H_{\text{L}}$ projection thus becomes an $X^{\otimes7}$ projection, that is, a non-destructive $X^{\otimes7}$ measurement.

\begin{figure}[t]
  \centering
  \includegraphics{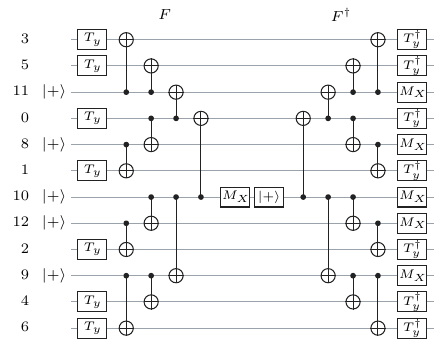}
  \caption{One cultivation round on the patch sites of Fig.~\ref{fig:steane-patch}, with the data qubits between the $T_y$ and $T_y^\dagger$ layers and routing ancillas 8--12 initialized in $\ket{+}$. The fold $F$ ends on root 10, which is read out in $X$ ($M_X$) and reset to $\ket{+}$ before the unfold $F^\dagger$. The final $X$ readouts of the routing ancillas give the second readout of root 10 and the flags of routing ancillas 8, 9, 11, and 12.}
  \label{fig:cultivation-round}
\end{figure}

We retain the double check~\cite{Gidney2024Cultivation, Rosenfeld2025MSC} but, unlike Ref.~\cite{Hirano2025MSCLS}, use ancilla 10 as the root rather than data qubit 0. Reading out a data-qubit root would destroy the encoding, so the second readout of the double check must first copy its value onto an ancilla. An ancilla root needs no copy, and both readouts, before the reset and after the unfold, measure it directly. In general, a nearest-neighbor C$X$ fold $F$ (Fig.~\ref{fig:cultivation-round}), with root $r$ and routing ancillas $A$ initialized in $\ket{+}$, satisfies
\begin{equation}
  F^\dagger X_rF=X_rX_AX_{\text{L}},\qquad X_A\equiv\prod_{h\in A}X_h.
  \label{eq:fold}
\end{equation}
The $+1$ root-measurement projector is $M_{r,+}=(I+X_r)/2$. On this branch the reset leaves the root unchanged, so the fold--measure--unfold Kraus operator is
\begin{equation}
  \begin{aligned}
    F^\dagger M_{r,+}F
    &=F^\dagger\left(\frac{I+X_r}{2}\right)F\\
    &=\frac{I+F^\dagger X_rF}{2}\\
    &=\frac{I+X_rX_AX_{\text{L}}}{2}.
  \end{aligned}
  \label{eq:fold-measure-unfold}
\end{equation}
Let $\ket{+}_{rA}=\ket{+}_r\bigotimes_{h\in A}\ket{+}_h$. Since $X_rX_A\ket{+}_{rA}=\ket{+}_{rA}$, the unnormalized accepted state for arbitrary data $\ket{\varphi}$ is
\begin{equation}
  \begin{aligned}
    &F^\dagger M_{r,+}F\,\ket{+}_{rA}\ket{\varphi}\\
    &\quad=\frac{I+X_rX_AX_{\text{L}}}{2}\,
      \ket{+}_{rA}\ket{\varphi}\\
    &\quad=\ket{+}_{rA}\,\frac{I+X_{\text{L}}}{2}\ket{\varphi}.
  \end{aligned}
  \label{eq:fold-on-state}
\end{equation}
Equivalently, the operator on the data is
\begin{equation}
  {}_{rA}\!\bra{+}F^\dagger M_{r,+}F\ket{+}_{rA}
  =\frac{I+X_{\text{L}}}{2}.
  \label{eq:fold-projector}
\end{equation}
Inserting this operator between the physical $T_y$ layers yields the logical Hadamard projector in Eq.~\eqref{eq:hadamard-projector}.

On the other branch, where the first root outcome is $-1$, the root is in $\ket{-}$. Its reset to $\ket{+}$ is equivalent to applying $Z_r$. This byproduct $Z_r$ stays on the root because $F^\dagger Z_rF=Z_r$, so its second $X$ readout reproduces $-1$.

\section{Lattice-surgery escape}
\label{si:escape}

Surface-code (target) data are prepared in $\ket{+}$; one complete stabilizer round establishes the encoded $\ket{+_{\text{L}}}$ state and its syndrome reference. In each of the three merge rounds, the merging stabilizers $Z_{1a}$, $Z_{35bc}$, and $Z_{de}$ are measured alongside the surface-code stabilizers that commute with them. They anticommute with the Steane stabilizer $X_{0145}$ and the surface code's boundary stabilizers $X_{ab}$ and $X_{cd}$. In the merged code, $X_{0145}$ and $X_{ab}$ survive only as their product $X_{0145ab}$, and $X_{cd}$ is dropped, so all three are withheld during the merge. We measure $Z_{35bc}$ in the order 3, $b$, 5, $c$, alternating between the codes, so that the hook error from a fault on its ancilla, $Z_{5c}$, places a single $Z$ on each code rather than both on the same code. Splitting it into $Z_{3b}$ and $Z_{5c}$ would avoid hook errors but anticommute with $X_{0235}$ and a surface-code stabilizer as well, changing more stabilizers of the merged code.

Our minimal Steane-check schedule measures $\{Z_{0145},Z_{0235}\}$ in the first merge round and $\{Z_{0145},Z_{0235},Z_{0246}\}$ in the second, with no Steane checks in the third. As in Sec.~\ref{si:superdense}, these checks keep their flags. At the split, the destructive transversal $X$ readout of the Steane data and one measurement of $X_{ab}$ reconstruct $X_{0235}$, $X_{0246}$, and $X_{0145ab}$. 

All Steane checks and flags are postselected. While the merging stabilizers may have either sign, a change between merge rounds causes rejection. All surface-code checks are decoded.

Viewed as teleportation, the merge and split leave the target in the source's input state up to a logical Pauli set by two outcomes (Table~\ref{tab:ls-feedforward}). The parity of the stable merging stabilizers $Z_{\text{L}}Z_{\text{L}}$ produces a logical $X$ feedforward and the Steane logical $X$ readout a logical $Z$ feedforward, both tracked classically and applied to the surface code. Full surface-code checks, including $X_{cd}$, then resume during post-transfer stabilization.

\begin{table}[t]
  \caption{Logical state of the target after lattice surgery, and the feedforward that returns it to the source's input $\ket{\psi_{\text{L}}}$, for each outcome $m_{ZZ}$ of $Z_{\text{L}}Z_{\text{L}}$ and $m_X$ of the source's $X_{\text{L}}$.}
  \label{tab:ls-feedforward}
  \begin{ruledtabular}
  \begin{tabular}{cccc}
    $m_{ZZ}$ & $m_X$ & Target state & Feedforward \\
    \hline
    $+1$ & $+1$ & $\ket{\psi_{\text{L}}}$ & $I$ \\
    $+1$ & $-1$ & $Z_{\text{L}}\ket{\psi_{\text{L}}}$ & $Z_{\text{L}}$ \\
    $-1$ & $+1$ & $X_{\text{L}}\ket{\psi_{\text{L}}}$ & $X_{\text{L}}$ \\
    $-1$ & $-1$ & $X_{\text{L}}Z_{\text{L}}\ket{\psi_{\text{L}}}$ & $Z_{\text{L}}X_{\text{L}}$ \\
  \end{tabular}
  \end{ruledtabular}
\end{table}

The complete fault-space search of Sec.~\ref{sec:msc-ls} uses the $S$ state as the Clifford proxy for the $T$ state, with a noiseless closing syndrome round and readout. Under full detector postselection, this minimal circuit has fault distance exactly three.

We provide here an account of why only $Z$ checks are needed during the merge. A $Z$ error on the Steane data commutes with the merging stabilizers, so it cannot change the $Z_{\text{L}}Z_{\text{L}}$ outcome, and the final transversal Steane $X$ readout, which reconstructs the Steane $X$ stabilizers, detects it. An $X$ error on data qubit 1, 3, or 5 flips $Z_{\text{L}}$ and a merging stabilizer but is invisible to that readout. If it enters during the merge, the merging stabilizer changes between rounds and the shot is rejected, but if it enters before the first merge round, it flips every round alike. Measuring $Z_{0145}$ and $Z_{0235}$, which together cover qubits 1, 3, and 5, in two rounds exposes such an error twice, so a single readout error cannot hide it. $Z_{0246}$ in the second round catches the pair $X_{05}$, which commutes with $Z_{0145}$ and $Z_{0235}$ but flips $Z_{\text{L}}$. Therefore, further removing any one of these five checks lowers the fault distance to two.

\section{Kickback tomography}
\label{si:kt}

Our injection and cultivation circuits are close to those of Ref.~\cite{Rosenfeld2025MSC} stage by stage (Table~\ref{tab:kt-circuits}). Both are executed on 14 sites, encode the injected state into the Steane code with a unitary encoder in nine layers of two-qubit gates, measure all six Steane stabilizers once after injection, and cultivate by a fold--measure--unfold measurement of $H_{\text{L}}$ with a double check and flags, and both preparation circuits have fault distance three. They differ only in the gate and measurement counts of each stage. In particular, we choose not to measure the encoder flags, which Ref.~\cite{Rosenfeld2025MSC} postselects, and our stabilizer round adds the flag of the copy of data qubit 2 (Sec.~\ref{si:superdense}).

\begin{table}[t]
  \caption{Two-qubit gates and measurements of each stage of the KT circuits of Ref.~\cite{Rosenfeld2025MSC}, counted from its released circuit, and of this work. A stabilizer round measures all six Steane stabilizers here.}
  \label{tab:kt-circuits}
  \begin{ruledtabular}
  \begin{tabular}{lcccc}
    Stage & \multicolumn{2}{c}{Two-qubit gates} & \multicolumn{2}{c}{Measurements} \\
    & Ref.~\cite{Rosenfeld2025MSC} & This work & Ref.~\cite{Rosenfeld2025MSC} & This work \\
    \hline
    Injection & 24 & 23 & 4 & 0 \\
    Stabilizer round & 30 & 34 & 6 & 7 \\
    Cultivation round & 24 & 22 & 6 & 6 \\
  \end{tabular}
  \end{ruledtabular}
\end{table}

\begin{figure}[t]
  \centering
  \includegraphics[width=0.8\columnwidth]{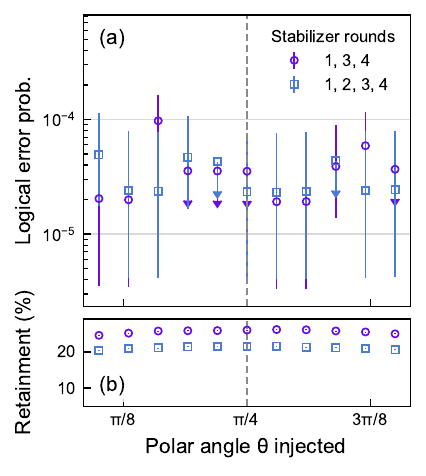}
  \caption{KT (a)~logical error probability and (b)~retainment against the injected polar angle $\theta$, with stabilizer rounds 1, 3, and 4 (circles) and with all four stabilizer rounds (squares), from $200{,}000$ shots per angle each. Error bars are 68\% confidence intervals. At angles where no logical error was sampled, the marker is the upper bound of the logical error probability and the arrow points toward zero. The dashed vertical line marks the $T$-state injection angle.}
  \label{fig:kt-check-rounds}
\end{figure}

Like Ref.~\cite{Rosenfeld2025MSC}, we chose where to insert Steane stabilizer rounds in the KT circuit by comparing options on hardware. The rounds can follow injection (round 1), the preparation round (round 2), and each of the two measurement rounds (rounds 3 and 4), before the destructive $X$ readout of the data. Ref.~\cite{Rosenfeld2025MSC} compared all four rounds with rounds 1 and 3 only and chose all four, which gave the lower KT logical error (Figs.~S18 and S19 of Ref.~\cite{Rosenfeld2025MSC}). We compared all four with rounds 1, 3, and 4 in one batch over the 11 dense angles (Fig.~\ref{fig:kt-check-rounds}). Without round 2, the logical error is unchanged within its uncertainty, while the retainment rises from $21.1\%$ to $25.5\%$, so we chose rounds 1, 3, and 4, which maintain fault distance three. Measuring and postselecting the encoder flags, tested in the same batch, likewise lowered the retainment without lowering the logical error. Fewer measurements mean fewer detectors that can reject a shot, which explains part of our higher retainment.

Each measurement round postselects its flags and the agreement of its two root readouts but keeps either root outcome. A logical error is a retained shot whose two root readouts both report the state orthogonal to $\ket{T}$. The data pool five executions on \texttt{ibm\_phoenix}, three over all 42 injection angles and two over the 11 dense angles, each pairing every cultivation circuit with its injection-only reference.

Logical-error uncertainties are 68\% binomial confidence intervals, which also reproduce the error bars of Fig.~3 of Ref.~\cite{Rosenfeld2025MSC}. Retainment error bars in the figures are 68\% binomial confidence intervals from pooled shots. The uncertainty of the $25.0(8)\%$ retainment at $\pi/4$ in Sec.~\ref{sec:results-kt} is instead the standard error of the mean over the five executions, because drift between executions far exceeds the shot noise. A maximum-likelihood fit of the injection-only error to $c+a\sin^2[(\theta-\theta_0)/2]$ (line in Fig.~\ref{fig:kt}) gives parameters very close to the ideal $c=0$, $a=1$, and $\theta_0=\pi/4$. The injection-only retainment does not depend on $\theta$, since its detectors do not depend on the logical state. At $\pi/4$, our KT logical error is $9.4$ times lower than the $1.9\times10^{-4}$ in the released data of Ref.~\cite{Rosenfeld2025MSC}, and our retainment is $2.9$ times their $8.7\%$.

\section{Comparison against previous grafting execution}
\label{si:willow}

Table~\ref{tab:willow} compares our execution of the grafting circuit with the Willow data of Ref.~\cite{Rosenfeld2025MSC}. Both are decoded by Tesseract with the search budget of Ref.~\cite{Rosenfeld2025MSC}, the Willow data with the SI1000 circuit-noise model~\cite{Gidney2022Honeycomb} at $p=10^{-3}$ and ours with the calibrated error model (Sec.~\ref{si:decoding}). Our execution is ahead on every quantity in the table. Without postselection on our inserted round after injection, which brings the circuit closer to theirs, the retainment rises further, with unchanged fidelity. With the larger priority-queue limit of our other results (Sec.~\ref{si:decoding}), $p_{\text{L}}$ is $2.19(3)\%$, and the other quantities in the table are unchanged within their uncertainties. The two executions differ in both the device and the transpilation, since Ref.~\cite{Rosenfeld2025MSC} used a simpler dynamical-decoupling sequence, no twirling, and its own reinforcement-learning control calibration, so we cannot attribute the improvement to the device alone. Twirling and dynamical decoupling are not expected to change the retainment significantly, however, so the higher retainment suggests that part of the improvement comes from the device.

\begin{table}[t]
  \caption{Grafting on Willow (Ref.~\cite{Rosenfeld2025MSC}, Fig.~4 and Fig.~S20b) and on \texttt{ibm\_phoenix} (this work). Both $p_{\text{L}}$ fits start at $N=2$.}
  \label{tab:willow}
  \begin{ruledtabular}
  \begin{tabular}{lcc}
    & Willow & \texttt{ibm\_phoenix} \\
    \hline
    Retainment, $N=2$ & $31.1\%$ & $44.3\%$ \\
    Retainment, $N=3$ & $37.0\%$ & $44.1\%$ \\
    Fidelity, $N=2$ & $0.954(2)$ & $0.9707(8)$ \\
    Fidelity, $N=3$ & $0.932(2)$ & $0.9505(8)$ \\
    Bloch radius, $N=2$ & $0.908(4)$ & $0.941(2)$ \\
    Bloch radius, $N=3$ & $0.864(4)$ & $0.901(2)$ \\
    $p_{\text{L}}$ per stabilizer round & $5.3\%$ & $2.28(3)\%$ \\
  \end{tabular}
  \end{ruledtabular}
\end{table}

\section{Layout selection}
\label{si:layout}

For each escape circuit, we select the layout once and apply it to every tomography basis and $N$. The search is exhaustive over native embeddings without SWAP gates. It first minimizes the number of couplers with a calibrated C$Z$ error above $5\%$ that the embedding uses, and then the cost $\sum_k w_k\,[-\ln(1-e_k)]$ over the two-qubit gates, physical single-qubit gates, and readouts of the circuit, where $e_k$ is the calibrated error of operation $k$ and $w_k$ counts how often the circuit applies it. Operations on qubits whose readouts are postselected weigh three times more, because a detection there discards the shot. All escape circuits use this procedure with the same settings, which gives the placements of Fig.~\ref{fig:device}.

\section{Circuit depth and retainment breakdown}
\label{si:stages}

Table~\ref{tab:stages} breaks the retainment of our executions at $N=3$ into postselection stages to support two retainment claims of Sec.~\ref{sec:results-escape}, that LS retains more shots than grafting before escape mainly because of qubit placement, and that the escape reverses this order because the LS escape postselects more. Pre-escape stages act on the Steane code before the merge of MSC-LS or the extension cycle of grafting. The table shows the matchable-switched grafting; grafting, a separate execution, has the same circuit up to the switch. The matchable-switched grafting's escape adds the two grafted cycles and the switch cycle (Sec.~\ref{sec:grafting}), which postselect nothing, so the two breakdowns agree to within about half a percentage point at every stage.

Table~\ref{tab:stage-depths} gives the two-qubit depth of each stage. The LS escape is deeper than the extension cycle of grafting, but its final-code cycle is three times shallower.

Before escape, LS retains $9.3$ percentage points more shots than the matchable-switched grafting while postselecting 10 detectors against 16. MSC-LS (full) (Sec.~\ref{si:comparisons}) adds 8 pre-escape detectors to our minimal circuit at a cost of only $4.6$ points, so at a similar cost per detector the 6 additional detectors of grafting would account for about 3 of the $9.3$ points. This suggests that placement contributes substantially to the rest, though the estimate does not isolate the placement effect. Two further observations support the placement effect. First, the calibrated-noise simulation (Sec.~\ref{si:simulation}), whose error model carries the calibrated errors of each placement, also retains more shots before escape for LS, by an even wider margin. Second, Fig.~\ref{fig:device} shows a likely source, a coupler in the grafting patch with a far higher C$Z$ error than its neighbors.

The escape reverses the order. The LS escape rejects more than half of the shots that reach it, spread over the merge checks, the Steane checks during the merge, and the Steane $X$ check at destructive readout, whereas the single extension cycle of grafting rejects about a quarter. The matchable-switched grafting therefore ends its escape with more shots despite starting it with fewer.

\begin{table*}[t]
  \caption{Postselection at $N=3$ for MSC-LS and matchable-switched grafting, in record order, with the postselected readouts and detectors, the fraction of arriving shots rejected, and the fraction of all shots retained after each stage.}
  \label{tab:stages}
  \begin{ruledtabular}
  \begin{tabular}{lcccc}
    Stage & Readouts & Detectors & Rejected of arriving (\%) & Retained after (\%) \\
    \hline
    \multicolumn{5}{l}{\textit{MSC-LS}} \\
    Post-injection Steane checks & 4 & 4 & 11.0 & 89.0 \\
    Cultivation & 6 & 6 & 21.2 & 70.2 \\
    \quad Pre-escape total & 10 & 10 & 29.8 & 70.2 \\
    Merge checks & 9 & 6 & 27.0 & 51.2 \\
    Steane checks during merge & 5 & 5 & 20.0 & 41.0 \\
    Steane $X$ checks at destructive readout & 8 & 3 & 24.3 & 31.0 \\
    Superdense extraction flags & 6 & 6 & 3.1 & 30.1 \\
    \quad Escape total & 28 & 20 & 57.1 & 30.1 \\
    \quad Total & 38 & 30 & 69.9 & 30.1 \\
    \hline
    \multicolumn{5}{l}{\textit{Matchable-switched grafting}} \\
    Injection flags & 4 & 4 & 5.2 & 94.8 \\
    Post-injection Steane round & 6 & 6 & 16.2 & 79.5 \\
    Cultivation & 6 & 6 & 23.3 & 60.9 \\
    \quad Pre-escape total & 16 & 16 & 39.1 & 60.9 \\
    Extension cycle & 8 & 6 & 28.6 & 43.5 \\
    Grafted cycles and switch cycle & 0 & 0 & 0.0 & 43.5 \\
    \quad Escape total & 8 & 6 & 28.6 & 43.5 \\
    \quad Total & 24 & 22 & 56.5 & 43.5 \\
  \end{tabular}
  \end{ruledtabular}
\end{table*}

\begin{table*}[t]
  \caption{Two-qubit depths, identical across both tomography bases and Pauli twirls at each measured $N$. These two-qubit depths are unchanged by transpilation. Gates from different stages can overlap, so depths need not add.}
  \label{tab:stage-depths}
  \begin{ruledtabular}
  \begin{tabular}{lccc}
    Stage & MSC-LS & Grafting & Matchable-switched grafting \\
    \hline
    Injection & 9 & 9 & 9 \\
    Post-injection Steane checks & 8 & 10 & 10 \\
    Cultivation & 8 & 8 & 8 \\
    \quad Pre-escape total & 24 & 27 & 27 \\
    Surface-code preparation & 4 & -- & -- \\
    Three merge rounds & 18 & -- & -- \\
    Steane readout and surface-boundary recovery & 2 & -- & -- \\
    Extension cycle & -- & 11 & 11 \\
    Two grafted cycles before the switch & -- & -- & 23 \\
    Switch cycle & -- & -- & 12 \\
    \quad Escape total & 21 & 11 & 44 \\
    \quad Total through escape & 42 & 38 & 71 \\
    One final-code cycle & 4 & 12 & 12 \\
    \hline
    \multicolumn{4}{l}{\textit{Complete circuit}} \\
    $N=3$ & 54 & 60 & 93 \\
    $N=10$ & 82 & 137 & 170 \\
  \end{tabular}
  \end{ruledtabular}
\end{table*}

\section{Further escape results}
\label{si:escape-results}

\begin{figure}[t]
  \centering
  \includegraphics[width=\columnwidth]{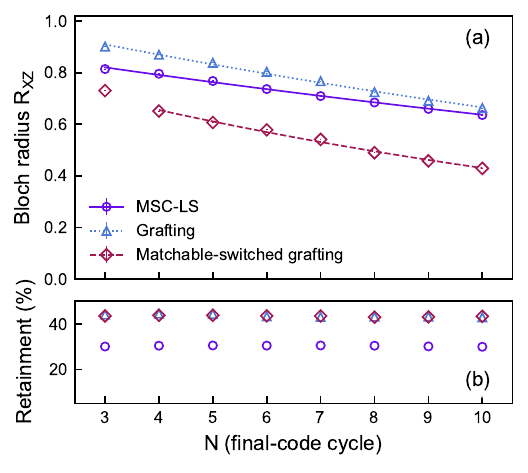}
  \caption{Without CGP, (a)~Bloch radius $R_{XZ}$ and (b)~retainment over both tomography bases against the final-code cycle $N$, for MSC-LS decoded with matching (circles), grafting decoded with Tesseract (triangles), and the matchable-switched grafting decoded with the hybrid decoder (diamonds). Lines are fits to $A\,e^{-2p_{\text{L}}N}$, which give $p_{\text{L}}=1.81(5)\%$, $2.23(4)\%$, and $3.50(7)\%$, respectively, with 95\% bootstrap intervals; the last fit starts at $N=4$.}
  \label{fig:decay}
\end{figure}

Figure~\ref{fig:decay} shows the $XZ$-plane Bloch radius and retainment against $N$ without CGP, with the fits that give $p_{\text{L}}$. The retainment barely changes with $N$, since the final-code checks are decoded rather than postselected.

The fit of $p_{\text{L}}$ (Sec.~\ref{sec:results-escape}) is unweighted least squares with $p_{\text{L}}\geq0$. Its 95\% interval comes from 10{,}000 bootstrap replicates of the whole sweep, each resampling all shots of every basis and $N$ before postselection and refitting $p_{\text{L}}$. Fidelity and Bloch-radius uncertainties are one standard error, propagated from the binomial variances of $\langle X_{\text{L}}\rangle$ and $\langle Z_{\text{L}}\rangle$.

\begin{figure}[t]
  \centering
  \includegraphics[width=\columnwidth]{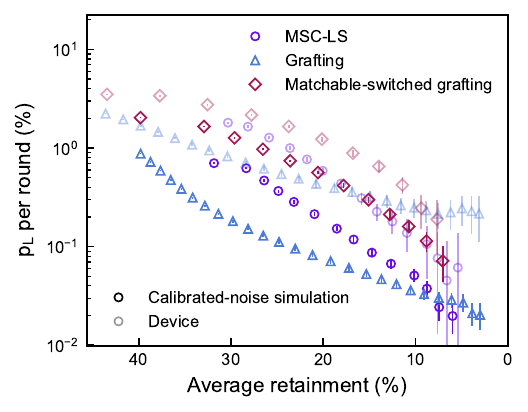}
  \caption{Calibrated-noise simulation (opaque markers) and device results (faded markers) of $p_{\text{L}}$ with 95\% bootstrap intervals against the average retainment over both tomography bases and the fitted $N$, from no CGP (leftmost) through rising CGP thresholds, for MSC-LS decoded with matching (circles), grafting decoded with Tesseract (triangles), and the matchable-switched grafting decoded with the hybrid decoder (diamonds). Simulated thresholds right after no CGP that do not lower $p_{\text{L}}$ are omitted.}
  \label{fig:si-simulation}
\end{figure}

\section{Decoding}
\label{si:decoding}

Every decoder uses a circuit-level error model of the submitted, transpiled circuit, one model per Pauli twirl. The non-Clifford Steane input, from injection through cultivation, is replaced by a Clifford proxy whose gates carry the calibrated noise of the physical operations they replace, and every later operation is converted with its calibrated noise. A calibrated average gate infidelity $r$ becomes a depolarizing channel of strength $3r/2$ after single-qubit gates and $5r/4$ after CZ gates, measurements flip at their readout error, resets (measurement feedforward) carry an error at the paired readout error, and idle periods carry the Pauli-twirled $T_1$ and $T_2$ channel. The model does not include leakage, crosstalk, coherent errors, correlated or asymmetric readout errors, or calibration drift.

Our LS circuit is decoded by minimum-weight perfect matching~\cite{HiggottGidney2025SparseBlossom}, correlated matching without CGP and plain matching with it, since the correlated reweighting distorts the complementary gap. Grafting is decoded by Tesseract~\cite{Beni2025Tesseract} with the search budget of Ref.~\cite{Rosenfeld2025MSC}, a detector beam of 10 with beam climbing, no revisited detectors, and six detector orders, except for a five times larger priority-queue limit of $50{,}000$, with which far fewer searches stop before the most likely error. The hybrid decoder of the matchable-switched grafting divides the history in time. Tesseract decodes its window, the detectors from the start of the circuit through the first two matchable cycles after the switch cycle, and commits only the faults it places through the switch cycle. It needs these two matchable cycles because an error near the switch is settled only after $d-1=2$ later comparisons for the fault distance $d=3$ that the circuit preserves. Matching decodes the detectors after the switch cycle, the first two matchable cycles included, with the committed faults as its time boundary.

\subsection{Complementary-gap postselection}
\label{si:cgp}

For CGP, every circuit-level model ends with one extra detector, the parity of the measurements that form the logical observable, whose ideal value is fixed. Setting its detection event to 0 or 1 while keeping all others forces the decoder to explain the shot without or with a logical flip, so each shot that passes the detector postselection is decoded twice, once in each logical sector. The lower-weight explanation gives the logical prediction. If the two weights are equal, we choose the sector without a logical flip. The complementary gap is the difference of the two weights. A weight is the sum of $\ln[(1-p)/p]$ over the faults of an explanation, each of probability $p$, so at a gap $g$ the more likely sector is about $e^{g}$ times as likely as the other. A threshold of 5, for example, retains only shots whose prediction is about 150 times as likely as its complement.

Matching decodes both logical sectors with plain minimum-weight perfect matching, and a shot that has no explanation in the opposite sector has an infinite gap. Tesseract decodes both sectors with the search budget of the ordinary decode, and a sector whose search ends without an explanation has infinite weight, so a shot with no explanation in either sector fails every threshold. The hybrid decoder needs the same comparison, but neither of its two decoders sees the whole history, so it assembles the most likely explanation in each sector from both. First, Tesseract decodes its window, the detectors through the first two matchable cycles after the switch cycle, twice. For this, the window's error model gets one more detector, flipped by every fault that flips both a detector in the window and the logical observable, which sets the window's logical sector just as the extra detector above sets that of the whole circuit. Each of the two window explanations gives one candidate set of faults to commit through the switch cycle. For each candidate, matching then decodes the remaining detectors in both sectors, starting from the committed faults as in the ordinary decode. The committed faults' weight plus matching's weight is the weight of a whole-history explanation, and in each sector the lighter of the two candidates' explanations is taken as its most likely explanation. The difference of these two sector weights is the complementary gap, so it reflects ambiguity inside the window as well as after it. Tesseract's window explanation also contains faults in the two matchable cycles after the switch cycle, but those are ignored after the commit, and matching explains the same cycles again, so adding them would count those cycles twice.

A threshold applies to both tomography bases and every $N$ of a sweep. The CGP sweeps of Figs.~\ref{fig:memory}(a) and~\ref{fig:fidelity} use thresholds of 0.5, 1, 1.5, and so on. The $p_{\text{L}}$ sweep requires at least $1\%$ retainment in each basis at every sampled $N$, whereas each fidelity panel requires it only at the $N$ it shows. Fidelity scans additionally cap the other protocols at grafting's last eligible threshold at that $N$. The highest observed fidelity at each $N$ (Fig.~\ref{fig:memory}(b)) needs no fit, so it is searched over a finer grid of thresholds spaced by 0.1, with the same fidelity-scan limits. The largest retainments at which each escape exceeds a fidelity of $0.99$ at $N=3$ (Sec.~\ref{sec:results-immediate}) are found on the same grid. We cap the fidelity at 1. At each threshold, $p_{\text{L}}$ is fitted to the retained shots at every fitted $N$, and the quoted retainment is the average over both bases and the same $N$. With CGP, the retainment falls with $N$ for every escape, by more than $40\%$ from $N=3$ to $10$ at a threshold of 5.

We observe that the CGP sweeps of LS and of the matchable-switched grafting end while their average retainment is still several percent. We first note that without CGP, the $X$- and $Z$-basis circuits of each escape retain the same fraction of shots. Then, with CGP, the $X$ basis of these two escapes loses shots faster, so their $X$-basis circuits at $N=9$ and $10$ are the first to drop below $1\%$ retainment, while grafting loses its two bases at similar rates. Under the $1\%$ requirement of the $p_{\text{L}}$ sweep, these $X$-basis circuits end the sweep while the $Z$ basis, and hence the average, still retains several percent. The $X$ basis loses shots faster because each basis is decoded from the checks that protect its observable, the $X$ checks for the $X$ basis, so noisier or sparser $X$ checks shrink the complementary gaps of the $X$ basis faster, even though both bases use the same qubits. In fact, both escapes have this problem. In LS they are noisier, because $X_{cd}$ is measured by the device's worst qubit and its detector fires about three times as often as the others, and in the matchable-switched grafting they are sparser, because the switch drops an $X$ stabilizer. The calibrated-noise simulation (Sec.~\ref{si:simulation}) shows the same asymmetry, which confirms that it comes from the circuits on their device layouts rather than from errors that the model omits. It is these noisier or sparser $X$ checks that end the two sweeps early.

Lastly, we note that decoding LS with Tesseract instead of matching gives a similar $p_{\text{L}}$ without CGP, but with CGP its $p_{\text{L}}$ stops falling near $0.35\%$ (Sec.~\ref{si:comparisons}). Tesseract's gap is search-budget-limited on shots with many detection events, so it lets some shots pass a CGP threshold that matching would reject. Since grafting is decoded with the same Tesseract, this suggests that the decoder contributes to the saturation of grafting's $p_{\text{L}}$ near $0.22\%$.

\begin{figure*}[t]
  \begin{minipage}[t]{\columnwidth}
    \centering
    \includegraphics[width=\columnwidth]{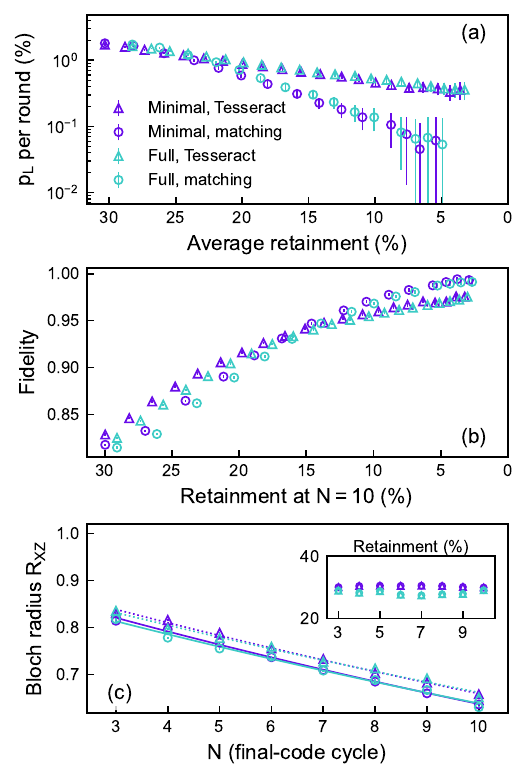}
    \caption{MSC-LS (minimal) and MSC-LS (full), each decoded with Tesseract and matching. (a)~$p_{\text{L}}$ against the average retainment over both tomography bases and the fitted $N$, (b)~fidelity with $\ket{T}$ against the retainment at $N=10$, each from no CGP (leftmost) through rising CGP thresholds, and (c)~Bloch radius $R_{XZ}$ without CGP against $N$, with the retainment in the inset. Error bars show $95\%$ bootstrap intervals in (a) and one standard error in (b) and (c).}
    \label{fig:si-ls-variants}
  \end{minipage}\hfill
  \begin{minipage}[t]{\columnwidth}
    \centering
    \includegraphics[width=\columnwidth]{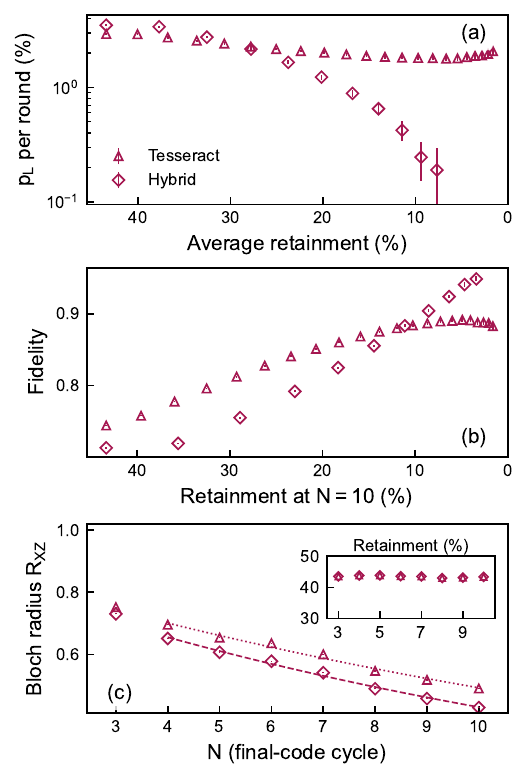}
    \caption{The matchable-switched grafting, decoded with whole-history Tesseract and with the hybrid decoder, in the panels of Fig.~\ref{fig:si-ls-variants}. }
    \label{fig:si-switch-decoders}
  \end{minipage}
\end{figure*}

\subsection{Cost}
\label{si:decoding-cost}

We timed each decoder in its own process on one node (AMD EPYC 9R45), on 500 shots of the $X$-basis circuit of one Pauli twirl at $N=10$, excluding the construction of the error model, and counted the ordinary decode and the two decodes of CGP together (Table~\ref{tab:decoder-cost}). Matching is at least three orders of magnitude faster than either decoder of the grafted codes. On the MSC-LS circuit itself, Tesseract is also more than three orders of magnitude slower than matching, so the speedup comes from the decoder rather than from the circuit. The hybrid decoder confines Tesseract to the history up to two rounds after the switch, so the Tesseract window has a fixed extent as matchable cycles are added, while the part that matching decodes grows; the grafted cycles before the switch still require Tesseract.

\begin{table}[t]
  \caption{Decoding time per shot at $N=10$, including the two decodes of CGP.}
  \label{tab:decoder-cost}
  \begin{ruledtabular}
  \begin{tabular}{lr}
    Decoder & Time (ms) \\
    \hline
    \multicolumn{2}{l}{\textit{MSC-LS}} \\
    Matching & $0.043$ \\
    Tesseract & $93.6$ \\
    \multicolumn{2}{l}{\textit{Matchable-switched grafting}} \\
    Hybrid & $158$ \\
    Whole-history Tesseract & $821$ \\
    \multicolumn{2}{l}{\textit{Grafting}} \\
    Tesseract & $393$ \\
  \end{tabular}
  \end{ruledtabular}
\end{table}

\section{Calibrated-noise simulation}
\label{si:simulation}

As a reference for the device results, we sample the error model of every submitted Pauli twirl of the three main sweeps, for as many shots as were executed on the device, and decode the samples with the device's decoders. The Steane input is a Clifford proxy (Sec.~\ref{si:decoding}), an eigenstate of the readout basis, so the $X$ and $Z$ circuits give the logical error probabilities $\varepsilon_X$ and $\varepsilon_Z$. Since the simulated circuit is Clifford with Pauli noise, its decoded logical noise is a Pauli channel, which shrinks each Bloch component by the same factor for every input state. A $\ket{T}$ state, with ideal $\langle X_{\text{L}}\rangle=\langle Z_{\text{L}}\rangle=1/\sqrt{2}$, therefore has $\langle X_{\text{L}}\rangle=(1-2\varepsilon_X)/\sqrt{2}$ and $\langle Z_{\text{L}}\rangle=(1-2\varepsilon_Z)/\sqrt{2}$, neglecting the difference between the $T$-state circuit and its Clifford proxy as in Sec.~\ref{sec:msc-ls}. From these expectation values, we compute the Bloch radius and fidelity and fit $p_{\text{L}}$ in the same way as for the device data.

Without CGP, the simulation reproduces the ordering of the three escapes on the device (Fig.~\ref{fig:si-simulation}). LS has the lowest $p_{\text{L}}$, grafting the highest fidelity at $N=3$, and both grafting escapes retain more shots than LS, which supports attributing these differences to the escape circuits on their device layouts. The simulated $p_{\text{L}}$ values are nonetheless $2.6$ times lower than measured for MSC-LS and grafting and $1.7$ times lower for the matchable-switched grafting, possibly due to errors that the model omits (Sec.~\ref{si:decoding}).

With CGP, the comparison changes. CGP lowers the $p_{\text{L}}$ of grafting more than that of MSC-LS in simulation, but less on the device, so at about equal retainment grafting with Tesseract reaches a slightly lower $p_{\text{L}}$ than MSC-LS with matching in simulation, the reverse of the device (Fig.~\ref{fig:si-simulation}). One possible explanation is the number of detection events per shot. Simulated shots likely carry fewer of them, possibly due to errors that the model omits, and on such shots Tesseract's gap is reliable, whereas device shots with many detection events can make it search-budget-limited (Sec.~\ref{si:decoding}); matching has no search budget to exhaust. Consistently, at a threshold of 5 the matchable-switched grafting, whose Tesseract decodes only a window, has a higher $p_{\text{L}}$ than grafting in simulation but a lower one on the device. The comparison of the main text is therefore made at the noise strength of current hardware, which the simulation does not reach.

\section{Supplementary analyses}
\label{si:comparisons}

We observe a rise in the fidelity of the matchable-switched grafting from $0.969(3)$ at $N=5$ to $0.976(3)$ at $N=6$ (Fig.~\ref{fig:memory}(b)), a difference of about two standard errors. It is not a property of the device shots, because whole-history Tesseract decodes the same shots to a fall, from $0.988(2)$ to $0.981(2)$. Nor is it a systematic effect of the hybrid decoder, because the calibrated-noise simulation decoded with it also falls, from $0.9865(6)$ to $0.9842(5)$. It is thus most likely a statistical fluctuation in the shots that the hybrid decoder's CGP retains at $N=6$.

Figure~\ref{fig:si-ls-variants} compares our minimal LS circuit with MSC-LS (full), the variant that adds encoder flags, all six Steane stabilizers instead of the minimal three in the post-injection round, and the original Steane checks during the merge rounds of Ref.~\cite{Hirano2025MSCLS}, each decoded with Tesseract and with matching. Without CGP, the two circuits have nearly the same $p_{\text{L}}$ with either decoder, and their highest fidelities at $N=10$ with matching agree within their uncertainties. With CGP, matching reaches a lower $p_{\text{L}}$ than Tesseract for both circuits. The minimal circuit thus performs like the full one while postselecting fewer detectors.

Figure~\ref{fig:si-switch-decoders} compares the hybrid decoder of the matchable-switched grafting with Tesseract decoding of its whole history. Without CGP, whole-history Tesseract gives a lower $p_{\text{L}}$ and a higher fidelity at $N=10$. With CGP, the hybrid decoder reaches the higher fidelity at $N=10$, and the $p_{\text{L}}$ of whole-history Tesseract stops falling near $1.8\%$, whereas the hybrid decoder reaches $0.2(1)\%$ at $7.7\%$ retainment. With CGP, the hybrid decoder is thus the better of the two on these longer shots, as well as the faster, possibly due to whole-history Tesseract's gap being search-budget-limited on shots with more detection events. Both fits use $N=4,\dots,10$, for the reason given in Sec.~\ref{sec:results-escape}.

\clearpage
\bibliography{refs}

\end{document}